\documentclass[12pt,preprint]{aastex}
\slugcomment{}
\shortauthors{Willams et al.}
\shorttitle{HI Observations of the Stephan's Quintet}
\newcommand{\kms}{km s$^{-1}$}
\newcommand{\Ha}{H$\alpha$}
\newcommand{\arcm}{$^{\prime}$}
\newcommand{\arcmper}{\ifmmode \rlap.{' }\else $\rlap{.}' $\fi}
\newcommand{\arcs}{$^{\prime\prime}$}
\newcommand{\arcsper}{\ifmmode \rlap.{'' }\else $\rlap{.}'' $\fi}         
\begin{document}

\title{The VLA HI Observations of the Stephan's Quintet (HCG~92)} 

\author{B.~A.~Williams}
\affil{University of Delaware, Department of Physics and Astronomy,
Newark, DE 19716, USA}
\email{baw@udel.edu}

\author{Min S. Yun}
\affil{University of Massachusetts, Department of Astronomy,
Amherst, MA 01003, USA}
\email{myun@astro.umass.edu}

\and 

\author{L. Verdes-Montenegro}
\affil{Instituto de Astrof\'{\i}sica de Andaluc\'{\i}a, CSIC,
     Apdo. 3004, 18080 Granada, Spain}
\email{lourdes@iaa.es}
\medskip

\begin{abstract}
Using the VLA\footnote{The National Radio Astronomy Observatory is a facility 
of the National Science Foundation operated under cooperative agreement by 
Associated Universities, Inc.}, we have made spectral-line and continuum observations of the neutral hydrogen
in the direction of the compact group of galaxies Stephan's Quintet.  The high-velocity clouds
between 5600 and 6600 \kms, the disk of the foreground galaxy, NGC 7320, at 800 \kms, the
extended continuum ridge near the center of the group, and 3 faint dwarf-like galaxies in the
surrounding field were imaged with C, CS, and D arrays.  Four of the HI clouds previously
detected are confirmed. The two largest HI features are coincident with and concentrated
mainly along separate large tidal tails that extend eastward. The most diffuse of the four 
clouds is resolved into two clumps, one coincide with tidal features south 
of NGC 7318a and the other devoid of any detectable
stellar or \Ha\ sources. The two compact clouds, along the same line of sight, have 
peak emission at luminous infrared and bright \Ha\ sources probably indicative of 
star-forming activity. The total amount of HI detected at high
redshifts is $\sim 10^{10}M_\odot$. As in previous HI studies of the group, no detectable 
emission was measured at the positions of any high-redshift galaxies so that any HI still 
bound to their disks must be less than $2.4 \times 10^{7}M_\odot$.

\end{abstract}


\section{Introduction}
\label{sec:intro}

Stephan's Quintet (Arp~319; ``SQ" hereafter) was the first compact
group to be discovered in the late 1800's, and it is probably the best
known and most studied dense group at all wavelengths.  The original
membership identified by \citet{s77} included five, bright,
relatively isolated galaxies (NGC 7317, 7318a\&b, 7319, and 7320) in a
tight configuration on the sky.  In addition to their close proximity,
three of the spiral members (NGC 7318a\&b, 7319) have peculiar or
highly distorted optical images that include faint wisps, filaments,
and tails.  Given its high surface brightness \citep[22.3 mag
arcsec$^{-2}$;][]{Hickson82} and surface density
enhancement \citep[593;][]{sul87}, it is no surprise to find 
the quintet included among
the Hickson groups \citep{Hickson82,Hickson93}. While the physical compactness
of most HCG's have been challenged 
\citep[see][and references therein]{Barns96}, 
the optical evidence of recent tidal interactions leaves
little doubt that some of the galaxies in the quintet are physically
close.  As is typical of Hickson groups \citep{sul97}, 
the quintet contains a
discordant-redshift member (NGC 7320).  A difference in redshift of
nearly 6000 \kms\ between NGC 7320 and the high-redshift members was
first reported by \citet{burbidge61} who concluded that NGC
7320 is either a member of the dense group or a foreground galaxy.
The discovery of the discordant redshift in SQ and
similar groups such as Seyfert's Sextet and VV172 \citep{VV59}
sparked much debate concerning the relationship between
the high- and low-redshift galaxies in compact configurations and
non-cosmological redshifts.

     Initially, 21-cm line studies of SQ were
undertaken mainly with the hope that the radio observations would
provide distances independent of redshift, thereby resolving both
questions about the group's membership and the existence of
non-cosmological redshifts. \citet{Allen70} was the first to detect HI
emission in the direction of the quintet.  This emission, detected at
800 \kms, was associated with NGC 7320.  Using the integral properties
derived from the HI profile and comparing them with similar properties
of field galaxies, \citet{Allen70} concluded that NGC 7320 is a dwarf
galaxy at a distance consistent with its redshift.  Because Allen's
study did not resolve the question of group membership,
\citet{Balkowski73} and \citet{Shostak74} were motivated to search
for HI emission from other quintet galaxies that possibly could.  Both
detected HI emission near 6600 \kms\ which they attributed to NGC
7319.  After applying the same method used by  \citet{Allen70}, they
derived very different distances. \citet{Balkowski73}  placed the
galaxy at the same distance as NGC 7320 while  \citet{Shostak74}
found it
more likely that NGC 7319 belongs at a distance commensurate with its
redshift.  The difference in their distances can be traced to the
calibration samples that each adopted.  The main weakness of this
method used to derive distances was the underlying assumption that the
galaxies in the quintet have normal HI properties for their
morphological types and luminosity.  Using the Westerbork Synthesis
Radio Telescope (WSRT), \citet{Allen80} would test this
assumption by providing new information on the distribution and
kinematics of the HI gas in the quintet.  Their observations supported
the assumption of normalcy in the location and motion of the
low-redshift HI but raised serious doubt about the use of HI profiles
as distance indicators when the data showed the gas at 6600 \kms\ in
an extended cloud (2\arcmper 5 $\times$ 4\arcm) 
not directly associated with any
of the members. 

     The discovery of an isolated cloud in SQ was
significant because it, alone, changed the focus of subsequent 21-cm
line studies.  HI observations more sensitive to extended emission
over larger regions of the quintet were undertaken in an effort to
understand this anomalous feature in the context of the group's origin
and evolution.  Sensitive HI searches that used the Arecibo telescope
found the presence of low level (8 mJy) broadband emission between
5600 and 6600 \kms\ \citep{s80} and two more features extended
and offset from the quintet galaxies \citep{Peterson80}.  With better
angular resolution and sensitivity than the old observations
\citep{Allen80}, \citet{Shostak84} used the WRST to locate and
confirm the three cloud features at 5700, 6000 and 6600 \kms\ that had
been previously detected.  Their synthesis observations showed the HI
largely outside the optical boundaries of the galaxies in the quintet.

 This present study of SQ is part of a broader
investigation of the distribution and kinematics of HI in the least
and most compact associations in the Hickson catalog 
\citep{will99a,will99b,vm00a,vm00b,vm01,hu00}
 in order to analyze the type and effect of
the interactions that are taking place. If physically
dense, Hickson groups are natural sites for studying tidal
interactions and its effects, galaxy formation and evolution via
merging processes, and the dynamical evolution of galaxy groups.
Spiral galaxies normally more extended in HI than in starlight are
susceptible to tides and direct collisions and ought to be more
sensitive to recent interactions.  High resolution imaging of the
neutral hydrogen can be used to distinguish between the real and
illusory compact groups and may provide a sensitive test of the
dynamical evolution of the physically dense groups.  
We have analyzed the spatial
distributions and kinematics of HI within a subset of 16 groups
and proposed an evolutionary scenario where  the amount of detected HI
would decrease further with evolution, either by continuous tidal
stripping and heating or by shocks produced when galaxies penetrate 
the group's core and interact violently with the interstellar or 
intergalactic HI in the groups \citep{vm01}.
SQ, amongst the densest Hickson groups, is a vital constituent 
in this scenario, hence its detailed study  would be
essential in characterizing the extreme stages in galaxy interactions
and group evolution.

     There is strong observational evidence that SQ is
a physical system.  The filaments, tails, and peculiar images of the
galaxies are suggestive of interactions.  The distribution of the HI
outside the galaxies is consistent with a history of collisional
encounters that may have stripped the spiral members of their HI disks
\citep{Allen80,Shostak84}. Why then observe SQ with the
VLA?  What more had we hoped to learn from HI synthesis observations?
The WRST \citep{Shostak84} integrated map gives us a clear sense of
how the gas between 5700 and 6600 \kms\ is distributed.  Given the
higher angular and velocity resolution of the VLA, we saw an
opportunity to improve the kinematical description of the HI gas
already detected outside of the galaxies.  With its higher sensitivity
to faint compact emission, the VLA could detect additional isolated HI
clouds within the quintet and possibly resolve any weak HI emission
associated with tidal features or still bound to the disks of the
individual galaxies.

\section{Observations and Data Reduction}
\label{sec:observations}

Observations of SQ were first made with the VLA in its
3 km (C) configuration on January 1, 1991 using all 27 telescopes.
The observational parameters are summarized in Table~\ref{table:obs}. 
Additional observations of the HCG~92A (centered at 782 \kms)
and the emission at 5700 \kms\ missed in the original observations
were made in the ``CS" configuration, which includes the 3 km
baselines and some extra short spacings \citep{r99}.
To improve sensitivity to extended, low surface brightness
sensitivity, additional D-array data were obtained in March 28, 
1999.  Shortest baselines at the shadowing limit of 25 meter
are present, and structures as large as 15\arcmin\ in size should
be visible in each channel maps.  
All of the data are calibrated following the standard
VLA calibration procedure in AIPS and imaged using IMAGR.
Absolute uncertainty in the resulting flux scaling is about 15\%,
and this is the formal uncertainty we quote for all physical
parameters derived from the flux density.
The synthesized beam is produced using a robust weight of $R=0.5$ is
19.4\arcs\  $\times$ 18.6\arcs .  The resulting spectral-line maps have a
velocity resolution of 21.5 \kms\ and an rms noise level of 0.21 mJy
(0.38 K). The 3$\sigma$ HI flux limit in each map is about 0.015 Jy
\kms, corresponding to an HI column density limit of $5\times 10^{19}$
atoms cm$^{-2}$.  At the adopted distance of the compact group (85
Mpc) and at the distance of HCG 92a (10 Mpc), the mass detection limit
is $2.4\times 10^7 M_\odot$ and $3.5\times 10^5 M_\odot$,
respectively.\footnote{The atomic gas mass has been calculated as 
$M_{HI} = 2.36 \times 10^5~D^2_{Mpc} \int S_v dv$, where 
$S_v dv$ is in Jy \kms.}

\section{Continuum Emission}
\label{sec:continuum}

A line-free continuum image constructed by averaging the 21 line-free
channels is shown in Figure~\ref{fig:cont}, and the main parameters
of the emission are given in Table~\ref{table:cont}.  This $15''$ resolution
image has an effective bandwidth of 2.05 MHz centered on an effective
frequency of 1416 MHz.  The rms noise level achieved in
Figure~\ref{fig:cont} is 0.10 mJy beam$^{-1}$ (0.3 K).  This continuum
image is a factor of 3 more sensitive in flux density and about 30
times more sensitive in surface brightness than the best continuum map
produced by \citet{vdHulst81}.

In its gross details, our continuum map is in good agreement with the
$6''$ resolution image generated by \citet{vdHulst81} and
with other published continuum maps
\citep{Allen72,Shostak84}. Besides detecting the two unresolved
northern sources and the extended linear source between NGC 7318b and
NGC 7319, an unresolved continuum source coincident with the nucleus
of NGC 7318a is also clearly detected.  A possible detection of a source
associated with the nucleus of NGC 7320 is also made
(Fig.~\ref{fig:cont}). More importantly, the new continuum image
delineates the diffuse emission more clearly than ever before.

The bright radio nucleus of NGC 7319 has a flux density of $27\pm4$
mJy, in good agreement with the results of \citet{vdHulst81} and \citet{Allen72}.  
Their high-resolution VLA data show a short (6\arcs) jet-like
feature emanating southwestward from the nuclear source in NGC 7319
\citep[see Fig.~3 of][]{vdHulst81}.  
The other unresolved source to the
northwest has no optical counterpart brighter than 29 mag~(arcsec)$^{-2}$
\citep[see Fig.~10][]{Arp73}, and it may be an unrelated background
source.  Its flux density is $10\pm1$ mJy. The radio source
coincident with the nucleus of NGC 7318a was only marginally detected
(at the $3\sigma$ level) by \citet{vdHulst81}.  We confirm
their detection of the nuclear radio source in NGC 7318a and find that
its flux density is $1.4\pm0.2$ mJy consistent with their measurement.
The new continuum image (Fig.~\ref{fig:cont}) also shows a 3.6$\sigma$
source nearly coincident with the nucleus of NGC 7320.

The high surface brightness sensitivity achieved by the new
observations reveals, for the first time, the faint structure in the
extended radio ridge between NGC 7318b and NGC 7319. The new continuum
image shows the smooth connection of the entire north-south ridge,
from the bright northern unresolved source to the elongated
north-south feature 1\arcmper 5 to the south.  An eastern extension
connecting to the nucleus of NGC 7319 and a northwestern extension
coincident with the string of \Ha\ emission \citep{pl99,Sulentic01}
north of NGC 7318b are present in the faint radio emission. The total
flux density associated with the extended radio ridge is $48\pm7$
mJy, which is at least 60\% smaller than that measured by 
\citet{vdHulst81}.  
The brightness distribution and the extent of the
radio ridge closely follow the extended X-ray emission imaged with
ROSAT \citep{Sulentic95,Pietsch97} and the extended \Ha\ emission
\citep{Ohyama98,Xu99}. By
itself, the radio continuum emission in the direction of the quintet
might be interpreted as a chance superposition of a background
core-jet source unrelated to the group; however, the spatial
coincidence of the radio ridge, \Ha\ emission, and X-ray emission
would be unusual, and difficult to explain in radio plumes and rarely
detected in radio core-jet sources. Furthermore, the radio continuum
emission is less likely to be a random background source since its
east-west extension also coincides with \Ha\ emission at the same
mean velocity as the HI gas in the group.

\section{HCG~92a (NGC~7320)}

The new VLA HI image of NGC~7320 shown in Figure~\ref{fig:H92AMOMNT}a
reveals an unperturbed normal HI disk with a small central depression
surrounded by an HI ring or a pair of tightly wound spiral arms.  The
total extent of the HI disk is 2\arcmper 5 $\times$ 1\arcmper 5 
(7.3 kpc$\times$ 4.4
kpc at D = 10 Mpc) comparable in size to the optical image. 
We plot in the upper left corner of  Figure~\ref{fig:H92AMOMNT}a
the integrated profile obtained by integrating the flux density 
in each channel map. The total
HI flux, $\int S_vdv$, recovered by the VLA is $8.25\pm0.05$ Jy \kms,
corresponding to a HI mass $M_{HI} = 1.95\pm 0.01 \times 10^8 M_\odot$. 
Although our
sensitivity is sufficient to detect clouds with an HI mass of at least
$3.5\times 10^5 M_\odot$, no other HI sources within the 15\arcmin\
(45 kpc) radius region mapped at frequencies near 1417 MHz were detected.
The HI properties of NGC 7320 are listed in Table~\ref{table:HI92a}.

The apparent blue magnitude of NGC 7320 is listed as 13.23 mag by 
\citet[][RC3]{dv91}.  Corrections for internal extinction
\citep{st81}, for galactic extinction as determined
\citet{fi81} who adapted from \citet{bh78}, 
and for the effect of redshift are 0.53, 0.39, and 0.075 mag,
respectively.  Application of the three corrections above yields a
blue absolute luminosity $L_B=2.0 \times 10^9L_\odot$ 
for this small, Sd galaxy.  Its M$_{HI}$/L$_B$, a measure 
of the HI gas content in solar units, is only $0.10$ and is more
characteristic of the HI content associated with early-type spiral
galaxies.  We derive an HI surface density $\sigma_{HI}$ =
$M_{HI}$/0.25$\pi$ a$_0^2$ = 8.0 $\pm$ 10$^{-4 }$g cm$^{-2}$ =
 3.8 M$_\odot$ pc$^{-2}$, where
a$_0$ is the corrected isophotal major diameter taken from RC3.  This
value is abnormally low when compared to the average HI surface
density of Sd galaxies in a sample of nearby galaxies
\citep{hr88} and in the larger sample of \citet{rh94}.
The total HI mass we derive is in good agreement with the previous
measurements \citep{Allen70,Gordon79,Allen80,Sulentic83},
and this low M$_{HI}$/L$_B$ ratio is not the result of HI
missed by the VLA observations.

The observed HI kinematics (Figure~\ref{fig:H92AMOMNT}b, 
Figure~\ref{fig:H92ACH}) is that of
normal disk rotation, without any signs of tidal distortions. The
derived rotation curve, shown in Figure~\ref{fig:N7320rc}, gradually
rises out to a radius of 60\arcs (3 kpc), where it peaks at a
value of $95\pm3$ \kms.  Both fits to the HI image and kinematics give
a position for the galaxy's center at $\alpha$(1950) = $22^h33^m45^s.8$,
$\delta$(1950) = 33$^\circ$ 41\arcm\ 21\arcsper5, a position angle 
of $-51^\circ
\pm1^\circ$ for the major axis, and a disk inclination of $i=48^\circ
\pm 2^\circ$. We find that the HI isophotal and dynamical centroids
agree within a few arcseconds of each other; therefore, previous
reports of a displacement by \citet{Allen80} and \citet{Sulentic83} 
are not supported by the new data.
 
Using the best-fitting parameters to the derived rotation curve in
Figure~\ref{fig:N7320rc}, we estimate a total mass $M_T$ =
(3/2)$^{3/n}$DV$^2_{max}$R$_{max}$/G = 2.2 $\times$ 
10$^{10}M_\odot$, based on \citet{br60}
 parametrization of the rotation curve, where V$_{max}$ is the
maximum rotational velocity, R$_{max}$ is the radius at which this
velocity first occurs, and $n$ is a parameter that describes the shape
of the derived rotation curve beyond R$_{max}$.  We assumed in the
calculation of total mass above, $n=1.3$ which corresponds to a flat
curve beyond 85$''$. With the values of M$_T$, M$_{HI}$, and L$_B$ given, we
obtain for the integral properties of NGC 7320 M$_T$/L$_B$ = 11 in solar
units and M$_{HI}$/M$_T$ = 0.01. The value of M$_T$/L$_B$ is reasonable
for its morphological type, but the ratio M$_{HI}$/M$_T$ is low  
\citep{rh94}.  It is clearly the HI integrated properties of NGC 7320
that are peculiar and not its kinematics.

\citet{materne74} were the first to provide evidence that NGC
7320 is probably a member of the stable group that includes the
supergiant galaxy NGC 7331. Projected less than 100 kpc away from the
center of this galaxy, NGC 7320 is probably a companion to (if not a
satellite of) this prominent spiral \citep{Tammann70}. The
low HI integrated properties measured for the dwarf galaxy are more
likely related to interactions with the halo of NGC~7331 rather than
interactions among the high-redshift members of SQ.  In
summary, our new HI observations firmly establish that NGC 7320 is a
relatively unperturbed, foreground galaxy projected against the
high-redshift members of the quintet.

\section{The Quartet Region}
\label{sec:quartet}

If our assumption about the distance of NGC~7320 is correct, then the
other four galaxies, NGC~ 7317, 7318a, 7318b, and 7319, form a bright 
{\it quartet} at an adopted distance of 85 Mpc.  Assuming a constant
mass-to-light ratio of $M_{tot}/L_B=7$ in solar units, 
we derived this distance from a
recessional motion of 6480 \kms\ (corrected for galactic rotation and
Virgocentric flow) for the quartet's center of mass and a Hubble
constant H$_0$ = 75 \kms\ Mpc$^{-1}$. The neutral hydrogen is detected 
between 5597 and 6755 \kms\ well displaced from the positions of 
the high-redshift galaxies (Figure~\ref{fig:tot}), a
result first reported by \citet{Allen80} and later confirmed
in greater detail by \citet{Shostak84}.  The redshift range would 
place the gas at the same distance as the quartet's center of mass. 
No emission brighter than
0.4 mJy beam$^{-1}$ is detected at the positions of any high-redshift
galaxies in the quintet including NGC 7320c, a likely 
fifth member located 3$'$ east of NGC~7319.
The HI emission concentrates  in three discrete velocity bands:
6475 -- 6755 \kms, 
5939 -- 6068 \kms,  
and 5597 -- 5789 \kms.  
It is mostly contained within five physically distinct components: 
a large curved feature  composed of two arc-like clouds 
(Arc-S and Arc-N) just east of the quartet and west of NGC 7320c, 
two kinematically distinct clouds (NW-LV, NW-HV) projected along the 
same direction north of NGC 7318a\&b, and
one diffuse feature (SW) between NGC~7318a\&b and NGC 7317.

\subsection{Distribution and kinematics of the HI gas}
\label{sec:HIdist}

\subsubsection{Arc Feature -- Large Tidal Tail(s)?}
\label{sec:Arc}

  The bulk of the HI (58\%) is detected in the large L-shaped
arc feature located about 3\arcm\ (75 kpc) southeast of the quartet's
center (see Fig.~\ref{fig:tot}).  
It extends over 5\arcm\ in size and has a mean radial 
velocity of 6610 \kms, which is about 130 \kms\ redshifted
with respect to the group's center of mass.  This feature
gives an impression of being a single long structure ($>125$ kpc),
possibly of tidal origin, but a careful examination of its
morphology, kinematics, and velocity dispersion 
suggests two
distinct features.  The clear morphological evidence for
two separate features is the two parallel tidal
tails visible on the northeastern side of NGC~7319 in deep
optical images (see our Fig.~\ref{fig:tot} and also 
Fig.~8 of \citep{Arp76}. 

The northern part of the HI cloud (Arc-N; Fig.~\ref{fig:arc}a) 
traces the optical tail emerging eastward from NGC~7319 toward 
NGC~7320c.   The curvature of the HI contours matches that of 
the bright optical tidal tail.  Peak emission
where the column density reaches $10^{21}$ atoms cm$^{-2}$ 
($\alpha$(1950) = $338^{\circ}.47$, 
$\delta$(1950)=$33^{\circ}.70$) is coincident with a
15$\mu$m star-forming region \citep[Source B of][]{Xu99} and 
a luminous \Ha\ emission object \citep[Source C1 of][]{Arp73}.
The linewidth associated with this region is larger than 
100 \kms, substantially larger than other parts
where the linewidth is barely resolved by the 21.5 \kms\
channels (Fig.~\ref{fig:if1ch}).
Faint emission associated with this feature continues to 
extend 2\arcm\ (50 kpc) northward (Fig.~\ref{fig:arc}a) into the 
region between NGC 7319 and NGC 7320c where \citet{Shostak84}
also detected extended HI at $\alpha$(1950) = 338$^{\circ}$.46, 
$\delta$(1950) = 33$^{\circ}$.73.

The southern part of this cloud (Arc-S; Fig.~\ref{fig:arc}b) 
is coincident with the fainter, less
prominent optical tail passing behind NGC~7320 and curving well beyond
its southern tip. Emission associated with this feature is diffuse and
ends sharply in the northwest where the continuum ridge begins. The HI
emission peak in the southern tail is coincident with the \Ha\
regions A12 and A14 in \citet{Arp73} while the HI peak at the southern
tip of NGC~7320 includes the \Ha\ sources A1, A2, A3, and A4
\citep{Arp73}.

Comparisons between the WSRT and VLA maps in the direction of 
NGC~7320 are severely limited by the lack of WSRT data across a third of
the galaxy's disk \citep[see Fig.~3 of][]{Shostak84}. We do not detect
any emission brighter than 0.44 ($2\sigma$) mJy beam$^{-1}$ in the
region just south of NGC~ 7319 and east of NGC~7318b (Figure~\ref{fig:if1mom}).  
Yet, the WSRT integrated map in this direction \citep[Fig.~3 of][]{Shostak84} clearly
reveals an extended condensation having a HI column density between
(1-4) $\times$ 10$^{20}$ atoms cm$^{-2}$.  Even though this reported surface
brightness is well above the VLA sensitivity limit(see 
\S~\ref{sec:HIprofiles}), it was not imaged in our VLA map nor in Fig. 6 
of \citet{Allen80}. As we discuss in section 5.2, we suspect that the difference
is related to the effect of residual radar interference and unstable levels in the
continuum emission that compromised the quality of the WSRT data. 

The intensity-weighted mean velocity map (Fig.~\ref{fig:if1mom}b) 
shows nearly constant line-of-sight velocity along the entire
length of Arc-N.  In contrast, the
velocity gradient observed along Arc-S is monotonic and
nearly linear, characteristic of a streaming motion typically
associated with tidal tails \citep[e.g.][]{hib95}.   The radial
velocity increases by 100  \kms\ over an angular distance of $2.'5$ (62
kpc) in a northwestern direction across this fainter tail.  \citet{Shostak84} did not 
detect this feature; instead, they report a similar
velocity gradient across Arc-N that the VLA does not confirm. 

All of the \Ha\ regions cited above
have been cataloged by \citet{Arp73}  as high-redshift objects, i.e.,
optical sources with radial velocities between 5700 and 6700 \kms. C1
has a redshift greater than 6500 \kms. The ionized gas has radial
motions consistent with that of the HI gas detected in the directions
of both optical tidal tails. \citet{Moles98} found no evidence of low
redshift \Ha\ sources in the southern optical tail and concluded
that it is a background feature to NGC~7320. Our HI data
confirm this finding.  The similarity between the shape of the
HI emission and the optical tails and the spatial correlation between
the HI emission peaks and high-redshift \Ha\ regions would
indicate that the gas in the tidal tails is associated with
the quartet rather than with the disk of NGC~7320.

\subsubsection{NW Features \label{sec:NW}}

HI emission north of NGC~7318 is detected in two compact clouds 
at two {\it distinct} velocities, projected along the same 
direction at $\alpha$(1950) = -1\arcmper 2, $\delta$(1950) 
= 0\arcmper 25, near the crossing point of the two 
tidal features (Fig.~\ref{fig:if1mom} and \ref{fig:if2mom}) north 
of NGC 7318a\&b.  
In the IF band centered at 1389.9 MHz (6595
 \kms), emission around this position is present in 12 velocity channels
(Fig.~\ref{fig:if1ch}) and contained within a 
slightly resolved feature, NW-HV (Fig.~\ref{fig:if2mom}b). 
This feature was first discovered with WSRT by \citet{Shostak84}. 
It shows a large velocity gradient along a position 
angle of $\sim$ 50$^\circ$ with the mean radial velocity 
increasing 60 \kms\ to the SW over an angular
distance of 0.7\arcm\ or 17.3 kpc (Fig.~\ref{fig:if1mom}b). 
In the higher IF band centered at 1392.5 MHz (6025  \kms), HI is
present in 7 velocity channels (Fig.~\ref{fig:if2ch}) and is
confined to the more extended cloud feature NW-LV. 
This low-velocity cloud was also originally detected in the WSRT data
and was dubbed ``West A" by \citet{Shostak84}.  Its velocity
gradient is similar in magnitude but nearly opposite in direction
to that of NW-HV, along a position angle of
40$^\circ$, over an angular distance of 1\arcm\ or 25 kpc 
(Fig.~\ref{fig:if2mom}b).

Detailed examinations of the channel maps in Figs.~\ref{fig:if1ch}
and \ref{fig:if2ch} reveal clear differences between the two
NW features in that the NW-LV feature is much more extended
spatially ($>30$ kpc), following along both optical tidal features,
while the NW-HV feature is far more compact.  
The HI emission peaks in NW-HV and NW-LV are coincident at the
resolution of these observations and lie in the same direction as
the luminous 15 $\mu$m source A of \citet{Xu99} and 
several bright \Ha\ sources 
that are probably indicative of star-forming activity embedded
in the densest regions of the gas. B10, a luminous \Ha\ source
cataloged by \citet{Arp73}, sits near this HI peak.  \citet{Moles98}
identify this same \Ha\ source as K3 and report a radial velocity
of 6680 \kms, in good agreement with that reported by \citet{pl99} for their source 6 (6670 \kms), and this \Ha\ 
source is associated with the NW-HV feature.  On the other hand,
B11 and B13 \citep{Arp73}, other \Ha\ sources
that align with the HI peak, appear to be related to the NW-LV
feature. These \Ha\ sources, later renamed K1 and K2 by \citet{Moles98},
have a redshift cz $\sim$ 6020 \kms\ \citep{Moles98} in good
agreement with the motion of the HI gas in NW-LV. Redshifts reported
by \citet{pl99} for  their \Ha\ sources 1--5 near the HI peak also have
radial motions between 5950 and 6005 \kms . Most of the \Ha\ sources with
measured velocities have radial motions very similar to the HI gas in NW-LV
which has the smaller velocity dispersion (Table~\ref{table:HI}). 
The mean radial velocity of the ionized gas is consistent with the HI motions in both of
the NW features  (Fig.~\ref{fig:if2mom}b and \ref{fig:if1mom}b) 
and suggests that the HI gas and the ionized gas are physically related and are part of the 
same large structure.

\subsubsection{Southwest Feature \label{sec:SW}}

The remaining HI detected toward the quintet resides in a 
diffuse cloud just south of NGC 7318a\&b (Fig.~\ref{fig:if3mom}).  
This feature was originally detected by  \citet{Shostak84}
who named it ``West B", and we identify it as the SW component 
in  Table~\ref{table:HI}.  
With the IF band centered at 1394.3
MHz (5597 \kms ), we succesfully detected emission in 10 velocity
channels (Fig.~\ref{fig:if3ch}). The systemic velocity of the 
SW feature is at 5700 \kms\ (Table~\ref{table:HI}) 
and is consistent with the value obtained by \citet{Shostak84}. 
Both the velocity integrated HI image (Fig.~\ref{fig:if3mom})
and the channel maps (Fig.~\ref{fig:if3ch}) suggest that this
feature consists of two large clumps.  The northern
clump spatially coincides with the tidal feature seen just
south of NGC~7318a and a second stream of tidal features and
\Ha\ sources seen further south.  The second clump is located
about $1'$ (25 kpc) south, and no stellar or \Ha\ emission
are associated with it.

The HI peaks of the SW feature have a much
lower column density than the peaks associated with the other three
major features (Table~\ref{table:HI}). As in the NW feature, the HI
peaks in the SW feature also appear to coincide closely with 
the \Ha\ objects B3 and B4 from \citet{Arp73} or 
15 and 14 in \citet{pl99}.  Taken on its own, 
the velocity field seen in Fig.~\ref{fig:if3mom}b is similar
to that of a distinct, rotating galactic disk.  The 
motions of the gas are not compatible
with those measured near the nucleus of NGC 7318a \citep{Moles98}.
Instead the gas velocity field is closer to the radial motions in 
NGC~7318b.  

The images shown in Figure \ref{fig:if3mom} and \ref{fig:if3ch}
provide greater detail in the
distribution and kinematics of the HI than is available in the
corresponding Westerbork maps  \citep{Shostak84} and reveal for the
first time that the SW component is clearly distinct and separate
from the other HI features. When taken together, the unrelated
velocity gradients across the low velocity features, i.e., NW-LV and
SW components (Table~\ref{table:HI}), do not support the \citet{Moles97}
model of a single rotating disk of HI that is gravitationally bound to
NGC 7318b.  It is clear from the VLA HI data that the lower-velocity
gas emanates from two structures that are spatially and kinematically
distinct.

\subsubsection{Summary of the HI Features}

The new VLA observations improve our 
understanding of the nature and extent of the atomic gas
in SQ. The distribution of the HI gas detected with the VLA
follows very closely the high surface brightness features present in
the WSRT data.  The large eastern cloud resolves
into 2 kinematically distinct features that coincide with the optical 
tidal tails that extend eastward toward NGC 7320c. It is now clear that the 
two low-velocity clouds in the western part of SQ are physically and
kinematically distinct from each other and the rest of the cloud features.
The presence of ionized gas traced in \Ha, coincident both in location 
and velocity, is ubiquitous for all HI features.  

All but one bright structure present in the WSRT data are confirmed. The one
exception, found just south of NGC~7819 in the WRST map, is puzzling. Given
the VLA's flux density and surface brightness sensitivities, this source
should have been confirmed at the $\sim$20$\sigma$ level.  
The WSRT data by \citet{Shostak84} was affected by radar
interference, and this feature may be a resulting artifact
(see below).

\subsection{Integrated Profiles of Distinct Regions in the Quartet}
\label{sec:HIprofiles}

In order to investigate the global properties of the components described
above, we produced integrated line profiles of the HI emission within the
four regions shown in Figure ~\ref{fig:tot}. The integrated profiles 
of NW-LV, NW-HV, SW, Arc-S and Arc-N, shown in Figure \ref{fig:spec},
have been reconstructed point by point from the individual channel maps
(Figs. \ref{fig:if1ch}, \ref{fig:if2ch} and \ref{fig:if3ch}). 
Table~\ref{table:HI} lists the relevant parameters of these line profiles 
along with the integrated fluxes determined directly from 
the HI spectra (Figure \ref{fig:spec}), the total hydrogen masses
for each feature, and the corresponding values from Table 2 of
\citet{Shostak84} for comparison.

The most peculiar HI profile shape obtained in the direction of 
Arc-S (see Fig.~\ref{fig:spec}), is remarkably similar to
the Arecibo HI profile measured by  \citet{Peterson80} in
their direction ``SSWW" just south of NGC 7318a\&b and west of NGC 7320.  
This asymmetric profile shape is generated by the high-velocity gas 
in the faint tail behind NGC 7320 where the motions are observed 
to be very systematic (Fig.~\ref{fig:arc}b). Since the
line shape cannot be attributed to irregularities in the velocity
field, this result suggests that there is more HI in the
southeastern end of the tidal tail than toward the northwestern edge
where the continuum ridge begins.  
In region Arc-N, the line profile (Fig.~\ref{fig:spec})
has its smallest width and most symmetric shape illustrating
how uniform the motions are over this extended region. These motions 
can be understood if we are observing a sheet-like distribution with 
its thin axis oriented mostly along the line of sight. Any large-scale 
motions, if present in this feature, would occur
mostly in the plane of the sky and not along the line of sight where 
random motions are expected to be small given the three-dimensional 
distribution of tidal features.

The broadest profile is measured in the direction of NW-HV, where 
strong \Ha\ and mid-IR sources \citep{Xu99} are also present near the end 
of a tidal feature that may belong to NGC 7318a \citep{pl99}.
There are fewer \Ha\ emission sources at the radial velocity of NW-HV than 
at the lower velocity of the NW-LV component. This combination
of high column density and large velocity width measured for NW-HV (Table~\ref{table:HI})
could be produced by a projected structure with significant bulk motion along 
the line of sight. Alternatively, NW-HV could be a self-gravitating complex 
star-forming region on its way to becoming a dwarf galaxy and the large-scale 
motions may be an indication of its rotation.  \citet{Zwicky56} was 
the first to introduce the idea that self-gravitating objects in tidal tails
could evolve into dwarf galaxies.  If NW-HV is a self-gravitating body, its 
dynamical mass is least $3.2\times 10^{10} M_\odot$, given its unknown inclination. 
The neutral hydrogen measured in the direction of NW-HV would then 
account for no more than 3\% of its total mass. 

To facilitate comparisons with published spectra obtained with 
single-dish and other aperture-synthesis observations, we generated 
a composite spectrum (Fig.~\ref{fig:spec})
of all the flux detected in the quintet (Fig~\ref{fig:tot}). 
This spectrum clearly shows, as expected, the three velocity components at 
6600, 6000, and 5700 \kms\ with 
three emission peaks at $\sim$55, $\sim$15, and $\sim$10 mJy, 
respectively. The distribution of 
the relative flux in Fig.~\ref{fig:spec} is nearly identical 
with those of spectra obtained with 
WSRT \citep{Shostak84}, Arecibo \citep{Gordon79,Shostak74,baw87}, 
Nancay \citep{Balkowski73}, and the 
NRAO 91-m \citep{baw87} telescopes except that detection of the 
5700 \kms\ component is at best marginal in most of the
published profiles. Only the composite spectra of \citet{Shostak84} and 
\citet{Peterson80} show all three velocity components. 
The integrated flux reported by \citet{Peterson80} and \citet{Shostak84} 
for the 6600 \kms\ velocity component is at least a factor
of 2 higher than the VLA measurement (Table ~\ref{table:HI}). The 6000 and 
5800 \kms\ velocity components have integrated
fluxes that are also lower than their corresponding WSRT or 
Arecibo values, but the difference is much
less than that measured for the high velocity gas. If we are to 
believe their results, then between 50 and
60\% of the flux is missing from the VLA observations; 
most of which can be attributed to the high-velocity gas.  

This large discrepancy cannot be accounted for by differences in
the sensitivity or in the minimum spatial frequencies sampled by the
two arrays. The WSRT high-resolution map has a synthesized beam (FWHP)
of $24''\times 47''$ and the brightness sensitivity in the
final channel maps is $6.8 \times 10^{-4}$ mJy arcsec$^{-2}$.
The corresponding brightness sensitivity in the
final maps of the VLA is $7.9 \times 10^{-4}$ mJy arcsec$^{-2}$.
The shortest baselines included in the VLA and WSRT observations
are both $\sim 30$ m; therefore, smoothly distributed emission
features with a size about 10\arcm\ and larger are missed by both
instruments and do not contribute to the flux.
Close inspection of the WSRT composite spectrum
\citep[Fig.~7a;][]{Shostak84} reveals a narrow high-velocity profile
on top a strongly curved baseline from 5300 to 7500 \kms,
in contrast to the very flat baselines present in the VLA
spectra (Fig.~\ref{fig:spec}).   Baseline problems seen in an
interferometeric spectrum usually indicate a calibration or imaging
problem with the data.
\citet{Shostak84} reported that 15 out of 63 total
channels in their data were corrupted by radar interference.
Although they claimed to have removed affected data, even low level
residual interference signal would affect the resulting
images, particularly at the largest spatial scales, resulting
in extended negative sources or displacement
of emission peaks (e.g. see \S~\ref{sec:Arc}).  This might
be enough to explain all observed discrepancies.

Over the last 30 years, SQ has been observed with at least five different 
radio telescopes, three of which are single-dish antennas. Comparisons of the
integrated flux measured with smaller single-dish antennas could
give some indication of how much flux, if any, was missed with the synthesis 
array. Figure \ref{fig:Gs} shows a plot of the integrated flux measurements of
the high velocity gas as a function of telescope beam. We have chosen 
to compare the high-velocity gas because it was detected in all the published HI 
observations and given its extended distribution would more likely be affected 
by the sensitivity limitations and beam attenuation. 

The beam areas plotted in Figure \ref{fig:Gs} span roughly 2 orders of magnitude and 
most of the integrated flux values lie between 3.2 and 6.5 Jy  \kms\ consistent 
with the VLA measurement given uncertainties in receiver noise, baseline 
fits to the underlying continuum emission, and flux scale calibrations. 
The only measurement in Figure \ref{fig:Gs} commensurate with the higher WSRT value 
is the one derived from an Arecibo composite spectrum (\citet{Peterson80}) 
generated from individual observations made over a grid centered on the group. 
Their low resolution contour map \citep[plate 1; Figure][]{Peterson80} shows HI
emission that extends over $\sim$ 8\arcm\ at an integrated 
brightness of 8 K  \kms\ beam$^{-1}$ or 0.94 Jy \kms\ beam$^{-1}$ 
and fills a region as large as 6.5 Arecibo beams.  None of the emission detected with the VLA or 
with WSRT is observed to be this extended; however, the Arecibo observations 
are more sensitive, i.e., $\sim 10^{-5}$ mJy arcsec$^{-2}$. If we assume that 
the Arecibo map is representative of the true source brightness, then a 
velocity width of 280 \kms\ \citep{Peterson80} would 
give an average brightness of $10^{-4}$ mJy arcsec$^{-2}$ for the most 
extended emission. This is well below the brightness sensitivity of both WSRT and the 
VLA.  Given similar reported values of the integrated flux, it is difficult 
to understand how WSRT could have detected the same faint emission present 
in the Arecibo map.

The most sensitive observations with the NRAO 91-m telescope (Fig.~\ref{fig:Gs}) were made by 
 \citet{baw87}; their HI spectrum has an rms noise per channel of 2.9 mJy 
which is roughly a factor of 10 higher than the noise in the Arecibo spectrum but the NRAO 91-m 
beam is a factor of 10 larger, resulting in a sensitivity 
$\sim 10^{-5}$ arcsec$^{-2}$. Because the NRAO 91-m and the Arecibo telescopes 
have similar sensitivities, the NRAO 91-m observations can provide an independent 
check of the larger Arecibo measurement of the integrated flux and 
a confirmation of the extended emission. The faint emission present in the Arecibo
map fills a region that is $\sim 70$\% of the NRAO 91-m half-power beam. 
\citet{baw87} report an integrated flux of 4.8 $\pm$ 0.3 Jy \kms\ which is consistent with 
the value measured by \citet{Shostak74} but less than half of the highest 
reported Arecibo measurement. 

The faint emission in the Arecibo map accounts for 56\% of the 
total integrated flux measured by \citet{Peterson80}. In order to avoid 
detection with the NRAO 91-m telescope, the faint emission would have to be 
distributed well outside of the HPBW which is inconsistent with the angular 
distribution of the high-velocity gas in both the WSRT and Arecibo maps.  
Although sensitive to the faint emission in the Arecibo map, the NRAO 91-m 
telescope is unable to confirm this detection and raises doubts about 
whether the 8 K \kms\ emission present in the Arecibo is real or an 
artifact of beam smearing. It is worth noting that emission as bright 
as 16 K \kms\ in the Arecibo map matches in angular extent the HI detected 
with both the VLA and WSRT and has an integrated flux of 4.8 Jy \kms\ 
consistent with the measurements made with both the VLA and the NRAO 91-m
telescope. 

In summary, the observed line widths of the components detected in 
the direction of the quintet are all very small less than 270 \kms. Their 
angular extent tends to correlate inversely with line width, i.e., the most 
compact component, NW-HV, has the largest line width, while the most extended
feature, Arc-N, has the smallest (Table ~\ref{table:HI}).  This correlation may be 
related to how dynamically evolved each feature is.  NW-HV is coincident 
with that of known \Ha\ emission regions or dwarf galaxy candidates
 \citep{Hunsberger96} and its larger line width may be an indication of a 
local density enhancement and a deeper gravitational potential. The 
integrated fluxes measured with the VLA are all lower than their corresponding
WSRT values. There is more than a factor of two difference between the two 
measurements of the high-velocity gas near 6600 \kms. This discrepancy cannot
be explained by differences in sensitivity or in the limitations on the
minimum spatial frequencies that can be mapped by each instrument. In contrast,
the VLA measurement of the total integrated flux is in good agreement with
single-dish measurements made with Arecibo, Nancay, and the 
NRAO 91-m telescopes.

\section{Neighboring Galaxies}
\label{sec:neighbors}

It is not surprising that we detected three other galaxies within the primary 
beam since SQ is not isolated, with NGC 7320c, its closest bright neighbor, less than 
5\arcm\ from the quartet's center. \citet{Tammann70} was one of the
first to point out that SQ is probably part of the Zwicky cluster
$22^h31^m.2$+37$^\circ$32\arcm\ \citep{Zwicky68}. The five anonymous galaxies detected by
\citet{Shostak84} in the surrounding field support the idea that 
SQ is embedded in a much larger aggregate of galaxies.
An integrated map of the HI between 5900 and 6750  \kms \ in the
surrounding region (Fig.~\ref{fig:satgen}) shows the position of the
three faint neighbors we detected.  
Two of the galaxies at $\alpha$(1950) = 
22$^h$33$^m$34$^s$, $\delta$(1950) = 
33$^\circ$ 53\arcm\ 53\arcs\ and at $\alpha$(1950) = 22$^h$34$^m$23$^s$, 
$\delta$(1950) = 33$^\circ$ 41\arcm 17\arcs\ had
already been detected by \citet{Shostak84} and named 
Anon.~No. 2, and No. 4, respectively, in their list of five.
This earlier detection of HI in Anon.~No.
2 was quite marginal but we are able to confirm HI emission 
in both these galaxies and add a newly detected neighbor 
Anon.~No. 6 at $\alpha$(1950) = $22^h34^m56^s$,
$\delta$(1950) = 33$^\circ 49'~25''$.            

The HI emission associated with Anon.~No.~2 is seen in just
three channels.  The emission peak coincides 
with the optical galaxy as shown
in the left part of Figure \ref{fig:satmom}a.
HI emission in Anon.~No. 2 is centered near 6380 \kms and has a peak 
HI column density of 6.3 $\times 10^{20}$ atoms cm$^{-2}$.
The observed 
velocity gradient across this galaxy is rather shallow
(Fig. \ref{fig:satmom}a, right side) but consistent with a rotating 
disk viewed nearly face-on. 
Anon.~No. 4 has HI emission that is quite symmetrical 
(Figure \ref{fig:satmom}b, left) and  rotation is evident in the
channel maps that clearly show the position of the HI centroid
displaced along the same position angle as that of the galaxy's
principal axis (Figure \ref{fig:satmom}b, right). This is consistent 
with motions generated by
an inclined rotating disk of gas.
As in Anon.~No. 2, the distribution of the HI in Anon.~No. 6, 
is asymmetric with respect to the center of the galaxy 
(Figure \ref{fig:satmom}c, left); however, the major axis of the HI emission
follows, rather constantly, the position angle of the optical
disk (Figure \ref{fig:satmom}c, right). The optical and HI properties of the faint galaxies detected in the field surrounding SQ are summarized in 
Table~\ref{table:sat}.

\section{Discussion} 
\label{sec:discussion}

New synthesis observations have motivated us to re-assess previously
published results and revise dated images that have characterized the
HI morphology of SQ for nearly twenty years.  The new observations at
higher angular and velocity resolutions than any published results to
date provide far more details about the internal motions of the HI gas
in SQ and allow direct comparisons between HI emission and other
discrete and extended sources within the group over a range of
wavelengths, i.e., X-ray, optical, \Ha, continuum, and infrared.

The higher resolution of the VLA has provided us
with additional clues about the short-term consequences of
collisions and interactions as well as the
dynamical history of the high-redshift
members in the quintet.  New details on substructures within all HI features 
detected, including the foreground galaxy NGC~7320, and the HI cloud at 6600 \kms, 
hold the valuable clues about the dynamical history of the group.

\subsection{Gaseous Collisions in the Group} 

The large velocity difference of NGC~7318b with respect to the
rest of the group ($\Delta V \sim 700$ \kms) has been taken as
key evidence for its being an intruder passing through the
group.  A collision at such a high velocity would result in
wide-spread shocked and ionized regions, followed by rapid cooling due to
the emission of radiation at UV- and X-ray wavelengths
\citep[e.g.][]{Harwit87}.
The radio continuum ridge found east of NGC~7318b has been
suggested to be the interface between the interloper and the
group because the shocked, ionized gas is predicted to
produce significant amounts of synchrotron emission \citep{vdHulst81,Allen72}.
The synchrotron life time for the radiating electrons is
estimated to be rather short, a few times $10^7$ years,
which is comparable to the crossing time for such a collision
\footnote{A classic example of enhanced radio synchrotron emission
arising from a collision involving two galaxies is the
``Taffy'' pair \citep[UGC~12914/5;][]{Condon93}.  The
HI observations of this colliding system should also
help in understanding the response of cold gas within the progenitor's
disks (see below).}.
Presence of extended, diffuse X-ray emission associated
with the group \citep{Sulentic95,Pietsch97} also raises the 
possibility that what is seen in SQ is the result of a collision
between a galaxy and a hot group medium.  Given the $n_e^2$
dependence on emissivity, such a collision should be much
less effective in producing radio continuum emission than
in a collision involving two galaxy disks.  A logical
place to look for such an example is the centers of nearby
clusters such as Virgo and the Coma cluster, and NGC~4522 may
be an example of such a collision between diffuse cluster/group
medium and a galactic disk \citep{Kenney99,Kenney00}. Though plausible,
such objects are rare in clusters, and the tenuous
intra-group medium in SQ would weaken the case considerably.

The excellent surface brightness sensitivity 
achieved by our new observations reveals several new details 
for the first time, including the faint structure in the
extended radio ridge between NGC~7318b and NGC~7319. 
As seen in Figure~\ref{fig:cont}, the continuum ridge 
approximately follows the optical tails. 
The new continuum image also raises minor issues for the 
hypothesis that the extended radio continuum ridge, also visible in 
X-ray and \Ha, is the product of a collision involving NGC~7318b.
Because NGC~7318b sits in the middle of the ridge its
trajectory has to be mainly in the
direction perpendicular to the extent of the ridge;
however, the shape of the radio ridge is S-shaped and not the expected
symmetric C-shape of a bow-shock.  

A careful comparison of radio continuum and \Ha\ 
images by \citet{Ohyama98} shown in Figure~\ref{fig:ridge}
reveals some important differences. The morphology of the \Ha\ emission
is indeed C-shaped, more in line with the shock hypothesis.
This apparent discrepancy may be resolved at least in
part by the difference in the radiation time scales.  The
cooling time for the \Ha\ emitting gas is much shorter than
the synchotron lifetime for the radio emitting electrons, and
\Ha\ is tracing only the most active, ongoing shocked regions.
There is also some evidence that more than one source of \Ha\ 
emission may be present.  The continuum
subtracted line image centered at 6738 \AA\ (including both
\Ha\ and [N II] emission) by \citet{Sulentic01} 
resembles our radio continuum image more closely,
and the radio continuum may be tracing mainly the ionized gas
at a the higher velocity (i.e. 6600 \kms) as shown in Figure 2 
of \citet{Xu99}.

The striking morphological alignment between \Ha\ and the
HI features near 6600 \kms\, shown on the right panel in
Figure~\ref{fig:ridge} raises an intriguing possibility
that NGC~7318b may be colliding with part of the HI tidal tail of
Arc-S that runs behind NGC~7320.  Implicit in this hypothesis is that
this HI tail may be linked with the NW-HV feature, and
the missing segment in HI may be fully ionized and visible
only in \Ha\ and in radio continuum.  The recession velocity
of the ionized gas in the southern parts of the \Ha\ filament
measured by \citet{Ohyama98} is around 6600 \kms, consistent
with this scenario.  Further, the broad linewidth 
($\Delta v \sim 900$ \kms) measured by \citet{Ohyama98} is 
also consistent with such a collision scenario.

\subsection{Star Formation Activities in SQ
\label{sec:SF} }

Examples of close correspondence between \Ha\ emission and HI are
found throughout SQ (see \S~\ref{sec:HIdist}).  The two prominent
star-forming complexes identified in mid-IR by \citet{Xu99}
are associated with two bright \Ha\ complexes, and they both
coincide with the two highest HI column density peaks (Arc-N and NW-HV).  
The HI peaks in the NW-HV, NW-LV, and the SW
features all have associated \Ha\ features, and the molecular
gas complexes traced in CO by \citet{Yun97} and by \citet{Gao00}
are also associated with \Ha\ features detected by
\citet{pl99}, \citet{Moles98}, \citet{Ohyama98}, 
\citet{Sulentic01}, and others.  

Frequent tidal
encounters in high density regions are thought to be conducive
to star-formation activity, and indeed evidence for induced
star-formation activity is ubiquitous everywhere in SQ where
cold gas is found.  The star-formation activity, however,
is not as spectacular as one might expect as in isolated environments
such as in interacting pairs.  
Star formation is known to have a non-linear dependence on 
gas density \citep[e.g.][]{Kennicutt98}, and few high density regions
are currently present in SQ and little cold gas is remaining in the 
individual galaxies.  These trends of low gas content
and ubiquitous but suppressed star-formation activity are 
commonly found in our examination of a large sample of
Hickson Compact Groups \citep{vm01}.

\subsection{Tidal Features and Interaction Scenarios} 

The improved sensitivity and resolution of the new VLA data
have significantly improved our understanding of the 
nature and distribution of the HI tidal features in SQ
(see \S~\ref{sec:HIdist}).   The massive L-shaped 
HI arc located about $3'$ (75 kpc) east of SQ is shown to
be possibly consisting of two separate tidal features (Arc-N and Arc-S in
Fig.~\ref{fig:arc}), in accordance with the two optical tails seen. Most of the gas in Arc-N
is located near the end of the long sinuous tail emanating from NGC 7319.
While Arc-N appears to originate from NGC~7319,
the origin of the southern tail is uncertain.  Both of the NW clouds
are located near the tips of 2 crossing optical features north of NGC 7318a\&b.
The two NW features are quite distinct in velocity, but they
appear spatially coincident exactly within our resolution.
Therefore the two may be physicaly related.  The SW feature
at 5700 \kms\ is shown to be a distinct feature consisting
of 2-3 clumps.  If only its velocity field is considered, 
it might be an independent dynamical entity, such as 
a dwarf galaxy (see \S~\ref{sec:SW}).

One of the clear conclusions that can be drawn from the new data is that
the 5700 \kms\ feature and 6000 \kms\ feature are clearly not the
tidally disrupted disk of NGC~7318b as proposed by \citet{Moles98}.
In fact, none of the HI features are clearly associated with any
of the individual galaxies.
It is conceivable that in the past the two features could have
been part of a single structure, possibly related to NGC~7318b,
and separated into two distinct features forming the components
we see today.  Such a separation cannot be a direct outcome of
a tidal disruption because {\it everything} within the tidal radius,
including gas and stars, should have responded identically to
the tidal shear.  To the contrary,
the core of the optical galaxy appears intact.  The suggestion
that NGC~7318b may be passing through the group for the first
time \citep{Moles98,Sulentic01} is certainly unlikely,
especially if the observed X-ray, \Ha, and radio continuum
ridge is the evidence for the ongoing collision, because
tidal damages due to such a collision, particularly the
removal of outer gas and stellar disk and the development of the
characteristic tidal tails, should form only after the
closest encounter.
Physical separation by a direct collision, though naively attractive,
is not a likely explanation because one cannot simply
slice a gas cloud into two pieces.
The face-on collision involving two disk galaxies in
the ``Taffy'' pair resulted in only a slight re-arrangement and
displacement of the HI in both disks \citep{Condon93}, without producing
any dramatic consequences of the type speculated here.

The general proximity and relative orientation of NGC~7320c with 
respect to the optical and HI tails have been invoked as evidence
for its recent passage through SQ and possibly as an explanation
for the northern optical tail that appears to originate from
NGC~7319 \citep[e.g.][]{Moles98,Sulentic01}.
\citet{Sulentic01} et al. even speculated that the two tidal tails might
be the products of two consecutive passages by NGC~7320c.
As clearly seen in various numerical studies of galaxy interactions
\citep[e.g.][]{BH96}, each passage by a companion generally 
erases all pre-existing features such as spiral arms and tidal tails,
and such an explanation for the pair of
tidal tails is probably not physical.  Another common tendency
seen in numerical simulations is that tidal tails generally 
points back to the responsible culprit because the tidal debris
and the passing companion have exchanged momentum, putting
them in similar trajectories about the primary.  This means
the northern HI tail, at least the bulk of the segment
pointing due north, {\it cannot be primarily driven by NGC~7320c}.
The origin for the older (more diffuse and thicker) 
southern tail is also quite uncertain.

The number of faint neighbors with similar radial velocities as the galaxies in SQ is evidence
that the quartet is not well isolated from the larger field of galaxies (see \S~\ref{sec:neighbors}).  Therefore
interlopers may pass through and disturb the
group with some frequency, including its HI companions such 
as Anon.~4.  Resulting dynamical heating may be sustaining
the group from collapsing \citep{Governato}, contrary to the prediction
by \citet{Barns96}.  One of the consequences of such a 
scenario is that interpreting the observed tidal features 
is no longer straightforward as in interacting galaxy pairs.
The high galaxy density in core of the group also means that
any galaxy passing through the group is subject to multiple
scatterings rather than following a simple trajectory, and
extrapolating the intuitions derived from the
numerical studies of interacting pairs should be exercised with
great caution.

An entirely different scenario one might consider,
for at least parts of the
observed HI features such as the large arc feature, is that
these are remnants of the primordial gas clouds where the
group has formed out of or tidal debris formed during
the initial formation of the group, similar to the Leo ring
found around the M96 group \citep{Schneider85,Rood85,Schneider89}.
The Leo Ring is a nearly complete ring of HI with a diameter of 
about 200 kpc, thought to be orbiting the galaxies M105 (E0) and 
NGC~3384 (SB0) with a period of $4\times 10^9$ years.
When examined at the same HI column density level of $9\times 10^{19}$
cm$^{-2}$ shown in Figure~\ref{fig:tot}, both the Leo Ring and
the HI arc feature are partial rings of HI of $\sim100$ kpc in
size with linear velocity gradient perpendicular to their lengths.
They both surround a group of galaxies interior to the ring,
and their systemic velocities match the galaxy group velocity.
One significant difference is that at least parts of the HI arc 
in SQ have stellar counterparts, which rejects the primordial
origin.  The possibility that these rings of HI are remnants
of group formation process, older than the group crossing time
still seems plausible.

\section{Summary}

Using C, CS, and D arrays of the VLA, we imaged five kinematically distinct HI clouds, a diffuse
extended continuum ridge, and the disks of NGC 7320, and 3 faint dwarf galaxies in the direction
of SQ.  The most massive of the four clouds previously detected and confirmed by us is resolved
into two separate systems which accounts for the five clouds we report here.  The two largest HI
features east of the continuum ridge are coincident with and concentrated along the two large tidal
tails that extend toward the direction of the NGC 7320c.  The two compact HI sources, west of
the continuum ridge, have HI emission that peaks along  the same line of sight and  includes
luminous infrared and bright \Ha\ sources probably indicative of star-formation activity in the
densest regions of the gas. The fifth cloud, south of NGC 7318a\&b, consists of two major clumps,
one coincides with the tidal features south of NGC 7318a and the other is devoid of any detectable
stellar or \Ha\ sources.  As in previous HI studies, no emission was detected at the positions of
any high-redshift galaxies including NGC 7320c; any HI still bound to their disks must be quite
small, i.e, less than $2.4 \times 10^{7}M_\odot$. 

We do not confirm the amount of HI previously reported in SQ.  The total amount of HI detected
by the VLA between 5597 and 6755 \kms\ is roughly $10^{10}M_\odot$ which is a factor of 2  lower
than that reportedly detected by WSRT.  Most of this difference can be attributed to the
integrated flux of the high velocity cloud features. The discrepancy cannot be explained by
significant differences in the instrumental performance of the two arrays and we suspect that it is
related to the quality and processing of the WSRT data which probably was compromised by low
residual radar interference and unstable levels in the continuum emission. We are more confident
in the VLA measurements because the data were free from these effects and the value of the
integrated flux of the high velocity gas is consistent with independent single-dish measurements
obtained with smaller antennas.  

The optical, X-ray, and radio continuum emission leaves little doubt that recent interactions have
occurred in SQ.  The radio continuum ridge visible in X-ray and \Ha\ is the strongest evidence of a
gaseous collision. But a collision between what two objects or components?  NGC 7318b has
been identified as the prime intruder and collision suspect because it brings enough kinetic energy
into the group to shock and ionize the gas. The continuum ridge in SQ, truly impressive in its scale,
is about a factor of 4 larger than the radio bridge between the "Taffy" pair and exceeds the
angular size of any of the bright galaxies in SQ. Given the large scale of the radio ridge, we favor
the hypothesis that the HI tidal tail behind NGC 7320 originally filled the region connecting Arc-S
and NW-HV and that a collision between NGC 7318b and this tidal tail fully ionized the HI
intragroup gas now detectable as the \Ha\, X-ray, and radio continuum ridge. The intruder theory
is attractive for three reasons: 1) there is a supply of faint HI-rich galaxies $\sim15'$ from SQ's
center; 2) the energy that the intruder deposits via interactions can be transformed into the
thermal and nonthermal forms we observed in SQ; and 3) this source of energy can support the
group against rapid (a few crossing times) collapse.
  
Unfortunately, there is no well-defined interaction scenario that can explain the HI tidal features in
SQ because none of the clouds are clearly linked with any of the bright galaxies. Some of the gas
in Arc-N is at the end of a tidal tail that appears to emanate from NGC 7319 but the original
source of HI in Arc-S is uncertain.  The two NW features both strategically located near the tips
of two tidal tails have systemic velocities very close to those of NGC 7318 a\&b and could be
physically associated with their northern tails as \citet{pl99} suggest. The SW feature is the only
cloud that has a projected separation that is close to the one bright galaxy sharing a similar
systemic velocity and this HI feature is found also at the end of a tidal tail/arm. Numerical
simulations of \citet{BH92} have shown that large pools of gas can develop at the
tip of tidal tails during a single encounter between two galaxies. When taken together, the
conspicuous sites of the HI clouds in SQ are consistent with the predicted location of tidally-liberated gas that has the potential to develop into dwarf galaxies. It is difficult to trace the
culprit(s) involved in the interactions when the galaxies have undergone multiple scatterings as a
result of their high number density.  We must be cautious about inferences based on the results of
numerical studies of interacting pairs when there are multiple interactions because these
interpretations are probably not completely valid in this case. 

Intergalactic HI clouds with masses greater than $10^{8}M_\odot$  are extremely rare as numerous null
searches \citep{FT81, LS79} in nearby groups of galaxies have revealed
and  yet, SQ contains multiple cloud features clearly distinct from the disk of the luminous
galaxies with no detectable levels of HI. Given the rarity of intergalactic HI clouds and the
evidence of interactions (two large optical tidal tails, diffuse light surrounding the group,  the
radio continuum ridge, HI-deficient galaxies), a natural explanation for the HI is that its origin is
tidal. The only other known group with as many cloud features as SQ is G11 \citep{dv75}
or GWa \citep{Materne78}, the nearby  compact group that contains the Leo HI cloud
complex. The general morphology of the two cloud complexes is similar in their relative
dimensions, clumpiness, and ring-like shape. But unlike the intergalactic HI ring in Leo where
there is no hint of diffuse stellar emission, the clouds in SQ are projected along the same
directions as the blue and red diffuse stellar emission \citep{Moles98} that surrounds the entire
group. In addition, the high density regions of the HI are decorated with numerous \Ha\ sources
indicative of the current star-forming activity. There are some clumps of HI devoid of diffuse
emission and \Ha\ sources so that we cannot dismiss the possibility that some of the gas in SQ is 
primordial and remnants of the group formation process.   The origin of the HI features in SQ 
still remains unresolved.

\acknowledgements

The authors gratefully acknowledge the helpful discussions with
A. del Olmo, C. Mendes de Oliveira, M. Moles, J. Perea, J. Sulentic, 
and C. Xu.  We also thank Y. Ohyama for providing us
with his published \Ha\ image for comparisons and the referee for valuable comments.  The data presented
here were obtained using the VLA.  The National Radio
Astronomy Observatory is a facility of the National Science Foundation
operated under cooperative agreement by Associated Universities, Inc.

\clearpage
\begin{figure}
\caption{Contour
map of the 21-cm continuum emission in HCG 92 overlapped on a
R-band image obtained by J. Sulentic at the 3.5m CAHA telescope.
The contours are $-$0.2, 0.2, 0.4, 0.8, 1.6, 3.2, 6.4, 13 
and 26 mJy/beam.  The synthesized beam, 15.4\arcs\ $\times$ 
14.8\arcs, is shown on the upper left corner.  Coordinate offsets
are with respect to the optical nucleus of NGC~7319 at
$\alpha(B1950)=22^h 33^m 46.^s1$, $\delta(B1950)=33^\circ 43'~00''$.
\label{fig:cont}}
\end{figure}
\clearpage
\begin{figure}[ht]
\caption{
(a) Map of the HI column density distribution in HCG 92a superimposed 
on the same R image shown in Figure~\ref{fig:cont}.
The contours are 9, 18, 36, 60, 89 and 134 $\times$ 10$^{19}$ 
at~cm$^{-2}$ and the beam size 19.4\arcs\  $\times$ 18.6\arcs.
The  HI flux density in mJy is plotted as a function of heliocentric velocity
in \kms\ in the upper left corner.
(b) Map of the first-order moment of the radial velocity field.
The numbers indicate heliocentric velocities in km~s$^{-1}$.  
\label{fig:H92AMOMNT} }
\end{figure}
\clearpage
\begin{figure}[p]
\caption{
Channel maps of the 21-cm line radiation of HCG~92a superimposed on the same R 
image as in Figure~\ref{fig:cont}. The Palomar Observatory Sky Survey (POSS) image has been used
for the eastern part of the field. The contours are $-$1.6, 1.6, 3.2, 6.6,
13, 20, 26, 33 $\times$ 10$^{18}$ atoms cm$^{-2}$.
\label{fig:H92ACH} }
\end{figure}
\clearpage
\begin{figure}[t]
\caption{
Rotation curve for NGC~7320 derived from the intensity weighted mean
velocity map (Fig.~\ref{fig:H92AMOMNT}b).  The solid line represents
an exponential rotation curve of the form $V_{rot}/V_{max}=1-e^{-R/R_{max}}$.
The best fit parameters are $V_{max} =$ 99 \kms\ and $R_{max}= 85''$.
\label{fig:N7320rc} }
\end{figure}
\clearpage
\begin{figure}[p]
\caption{Map of the total 
HI column density distribution in HCG 92 superimposed on the same R 
image as in Figure~\ref{fig:cont}. The POSS image has been used
for the eastern part of the field. The integration
range is from 5597 to 6918 \kms\ and the contours
are 5.8, 15, 23, 32, 44, 61, 87, 120, 180
$\times$ 10$^{19}$~atoms cm$^{-2}$. The synthesized beam is 19.4\arcs\ 
$\times$ 18.6\arcs.
\label{fig:tot}}
\end{figure}
\clearpage
\begin{figure}[p]
\epsscale{0.7}
\caption{Map of the HI column density distribution in HCG 92 
Arc-N (a) and Arc-S (b)
superimposed in the same R image shown in Figure~\ref{fig:cont}.
The POSS image has been used for the eastern part of the field.
The contours are 1.5, 5.8, 12, 18, 23, 29, 44, 58, 73 and 87
$\times$ 10$^{19}$ atoms~cm$^{-2}$ and the synthesized beam is
19.4\arcs\  $\times$ 18.6\arcs. 
\label{fig:arc}}
\end{figure}
\clearpage
\begin{figure}[p]
\epsscale{0.8}
\caption{Channel maps of the 21-cm line radiation of HCG 
92 corresponding to Figure~\ref{fig:if1mom} superimposed on the same R 
image as in Figure~\ref{fig:cont}. The POSS image has been used
for the eastern part of the field. The contours are -3.4, 3.4, 
6.9, 10, 14, 20, 27 and 34 $\times$ 10$^{18}$ atoms cm$^{-2}$.
\label{fig:if1ch}}
\end{figure}
\clearpage
\begin{figure}[ht]
\caption{(a) Map of the HI column density distribution in HCG 92 
in the velocity range 6475 -- 6755 \kms\ superimposed 
in the same R image shown in Figure~\ref{fig:cont}. 
The POSS image has been used for the eastern part of the field.
The contours are  5.8, 12, 18, 23, 29, 44, 58, 
73 and 87 $\times$ 10$^{19}$ 
atoms~cm$^{-2}$ and the synthesized beam is 19.4\arcs\  $\times$ 18.6\arcs . 
(b) Map of the first-order moment of the radial velocity field.
The numbers indicate heliocentric velocities in km~s$^{-1}$.  
\label{fig:if1mom}}
\end{figure}
\clearpage
\begin{figure}[ht]
\caption{(a) Map of the HI column density distribution in HCG 92
in the velocity range 5959 -- 6068 \kms\ superimposed 
in the same R image shown in Figure~\ref{fig:cont}.
The POSS image has been used for the eastern part of the field.
The contours are 5.8, 12, 18, 23, 29, 44, 58 and 73 
 $\times$ 10$^{19}$ 
atoms~cm$^{-2}$ and the synthesized beam is 19.4\arcs\  $\times$ 18.6\arcs . 
(b) Map of the first-order moment of the radial velocity field
in the same velocity range as in Figure~\ref{fig:if2mom}a.
The numbers indicate heliocentric velocities in km~s$^{-1}$.  
\label{fig:if2mom}}
\end{figure}
\clearpage
\begin{figure}[p]
\caption{Channel maps of the 21-cm line radiation of HCG 
92 corresponding to Figure~\ref{fig:if2mom} superimposed on the same R 
image as in Figure~\ref{fig:cont}. The POSS image has been used
for the eastern part of the field. The contours are 
$-$3.4, 3.4, 6.9, 10 and 14 $\times$ 10$^{18}$ atoms cm$^{-2}$.
\label{fig:if2ch}}
\end{figure}
\clearpage
\begin{figure}[ht]
\caption{(a) Map of the HI column density distribution in HCG 92
in the velocity range 5597 -- 5789 \kms\  superimposed 
in the same R image shown in Figure~\ref{fig:cont}.
The POSS image has been used for the eastern part of the field.
The contours are 5.8, 15, 23, 32 and 41 $\times$ 10$^{19}$ 
atoms~cm$^{-2}$ and the synthesized beam is 19.4\arcs\  $\times$ 18.6\arcs . 
(b) Map of the first-order moment of the radial velocity field
in the same velocity range as in Figure~\ref{fig:if3mom}a.
The numbers indicate heliocentric velocities in km~s$^{-1}$.  
\label{fig:if3mom}}
\end{figure}
\clearpage
\begin{figure}[p]
\caption{Channel maps of the 21-cm line radiation of HCG 
92 corresponding to Figure~\ref{fig:if3mom} superimposed on the same R 
image as in Figure~\ref{fig:cont}. The POSS image has been used
for the eastern part of the field. The contours are 
-3.4, 3.4, 6.9, 10 and 14 $\times$ 10$^{18}$ atoms cm$^{-2}$.
\label{fig:if3ch}}
\end{figure}
\clearpage
\begin{figure}[p]
\epsscale{0.50}
\caption{HI flux density in units of mJy as a function of heliocentric velocity in units of \kms\
for the features indicated in each plot. \label{fig:spec}}
\end{figure}
\clearpage
\begin{figure}[t]
\epsscale{1.2}
\caption{VLA and published HI Integrated fluxes of the high velocity gas between 6475 and 6755
\kms\ as a function of telescope beam area in units of arcmin$^{-2}$
on a logarithmic scale. Values measured with Arecibo, NRAO 91-m, 
Nancay, WSRT, and the VLA radio telescopes are labeled for comparison. \label{fig:Gs}}
\end{figure}
\clearpage
\begin{figure}[p]
\epsscale{1}
\caption{Map of the HI column density distribution in HCG 92 superimposed
in the same R image shown in Figure~\ref{fig:cont} for the central part and
the POSS image for the rest of the field.
The contours are 5.8, 13, 22, 35, 61, 87, 117 and 150 $\times$ 10$^{19}$
atoms~cm$^{-2}$.
The systemic velocities of the HI features (Table~\ref{table:HI}), the satellite galaxies (Table~\ref{table:sat}), and the bright galaxies \citep{Hickson92, Sulentic01} are labeled.
\label{fig:satgen}}
\end{figure}
\clearpage
\begin{figure}[p]
\epsscale{.65}
\caption{Map of the HI column density distribution (left) and
first-order moment of the radial velocity field (right)
of satellite galaxies (a) Anon.~2 (b) Anon. 4 and (c) Anon. 6
superimposed in the POSS image.
The column density contours are (a) 15, 26, 37, 48 and 59 $\times$ 10$^{19}$ 
atoms~cm$^{-2}$ (b) 7, 13, 20, 26 and 33 $\times$ 10$^{19}$ 
atoms~cm$^{-2}$ and (c) 4, 13, 22, 31 and 40 $\times$ 10$^{19}$ 
atoms~cm$^{-2}$. The synthesized 
beam is 17.5\arcs\ $\times$ 16.8\arcs. The numbers in the radial velocity field
indicate heliocentric velocities in \kms\ and the velocity contours are 
in units of 10 \kms. \label{fig:satmom}}
\end{figure}

\clearpage
\begin{figure}[ht]
\epsscale{1.1}
\caption{(a) A comparison of the radio continuum (in contours)
with the \Ha\ image by \citet{Ohyama98}.  The contour levels are
identical to those in Fig.~\ref{fig:cont}. (b) A comparison of
the HI column density distribution in HCG 92
in the velocity range 6475 -- 6755 \kms\ (in contours) 
with the same \Ha\ image.  The contour levels are 10\% of
the peak.
\label{fig:ridge}}
\end{figure}

\clearpage
\begin{deluxetable}{cccccc}
\tablewidth{0pt}
\tablecaption{Summary of Observations\label{table:obs}}
\tablehead{
\colhead{}              & 
\multicolumn{2}{c}{RMS Noise} &
\colhead{Channel Width}      & 
\colhead{Velocity Range}  &
\colhead{Beam Size}     \\
\colhead{} & 
\colhead{(mJy beam$^{-1}$)} & 
\colhead{(K)} & 
\colhead{(\kms)} & 
\colhead{(\kms)} &
\colhead{}
}
\startdata
Natural Weighting:&&&&&\\ 
H92a & 0.20 & 0.34 & 20.7 & 475 - 1097 & 19\arcsper 4 $\times$ 18\arcsper 6\\
IF1 & 0.21 & 0.38 & 21.5 & 6272 - 6918 & 19\arcsper 4 $\times$ 18\arcsper 6\\
IF2 & 0.21 & 0.38 & 21.5 & 5725 - 6326 & 19\arcsper 4 $\times$ 18\arcsper 6\\
\enddata
\end{deluxetable}

\clearpage
\begin{deluxetable}{lccc}
\tablewidth{0pt}
\tablecaption{1.4 GHz Continuum Emission in HCG~92 \label{table:cont}}
\tablehead{
\colhead{Name}      &
\colhead{Size}      & \colhead{Flux}  & \colhead{$L_{1.4GHz}$} \\
\colhead{}      &
\colhead{}      & \colhead{(mJy)}  & \colhead{(W~Hz$^{-1}$)} }  
\startdata
NGC~7317  & -- & $<0.3$ mJy & $< 2.3\times 10^{17}$ \\
NGC~7318a & $<5''$ & $1.4\pm0.2$ mJy & $(1.1\pm0.2)\times 10^{18}$\\
NGC~7318b & -- & -- & -- \\
NGC~7319  & $<5''$ & $27\pm4$ mJy & $(2.1\pm0.3)\times 10^{19}$ \\
NGC~7320  & $<5''$ &  $0.4\pm0.1$ mJy & $(4.8\pm1.2)\times 10^{15}$ \\
NGC~7320C  & $<5''$ &  $<0.3$ mJy & $< 2.3\times 10^{17}$ \\
Radio Core & $<5''$ & $10\pm1$ mJy & $(7.5\pm1.1)\times
10^{18}$\tablenotemark{\dagger} \\
Radio Filament & $0.5'\times 2.5'$ & $48\pm7$ mJy & 
$(3.6\pm0.5) \times 10^{19}$ \\
Total  & -- &  $96\pm 15$ mJy & -- \\
\enddata
\tablenotetext{\dagger}{Assuming a distance of 85 Mpc, the same 
as for SQ.}
\end{deluxetable}

\clearpage
\begin{deluxetable}{lc}
\tablewidth{200pt}
\tablecaption{Summary of HI Observations of HCG 92a\tablenotemark{\dagger}
\label{table:HI92a}}
\tablehead{
\colhead{} & \colhead{HCG~92A}  }
\startdata
$<V_{HI}>$(\kms)     & 776$\pm$10 \\
Velocity Width (\kms)& 192$\pm$10 \\
Size~(arcmin) & $2.5'\times 1.5'$ \\
~~~Size~(kpc)  & $7.3\times 4.4$ \\
Peak Flux (Jy  \kms) & 0.55$\pm$0.01 \\
~~~Peak $N_{HI}$ (10$^{21}$ cm$^{-2}$) & 1.65$\pm$0.03 \\
~~~Peak $A_V$ (mag) & 0.77  \\
\sidehead{ Integrated Flux (Jy  \kms):} 
~~~Shostak et al. (1984) & 7.8 \\
~~~New VLA & 8.25$\pm$0.05  \\
$M_{HI}$ ($10^9 M_\odot$) & 0.195$\pm$0.001 \\
\enddata
\tablenotetext{\dagger} {Assuming distance of 10 Mpc for HCG~92A.}
\end{deluxetable}

\clearpage
\begin{deluxetable}{lccccc}
\tablecaption{HI Parameters\tablenotemark{\dagger} of the 
Kinematical Components
\label{table:HI}}
\tablehead{
\colhead{} & \colhead{SW} & 
\colhead{NW-LV}  & \colhead{NW-HV}  &  \colhead{Arc-S} &  \colhead{Arc-N} }
\startdata
$<V_{HI}>$(\kms)     & $5699\pm15$& 6012$\pm$10& $6647\pm15$& $6641\pm10$& 6604$\pm$10\\
Velocity Width (\kms)& 213$\pm$5 &  160$\pm$5 &  255$\pm$5&$153\pm5$ &  114$\pm$5 \\
Size~(arcmin) & $1.5' \times 0.9'$ & $1.2'\times 1.0'$ & $0.7\times 0.5$ &
$3.1'\times 0.7'$ & $3.6'\times 1.3'$ \\
~~~Size~(kpc)  & $37\times 22$ & $30\times 25$ & $17\times 12$ &
$77\times 17$ & $89\times 32$ \\
Peak Flux (Jy  \kms) & 0.14$\pm$0.01 & 0.26$\pm$0.01 & 0.31$\pm$0.01 &
0.11$\pm$0.01 & 0.33$\pm$0.01 \\
~~~Peak $N_{HI}$ (10$^{21}$ cm$^{-2}$) & 0.41$\pm$0.03 &
0.76$\pm$0.03 & 0.90$\pm$0.03 & 0.32$\pm$0.03 & 0.96$\pm$0.03 \\
~~~Peak $A_V$ (mag) & 0.20  & 0.36  & 0.46 & 0.46  & 0.25   \\
\sidehead{ Integrated Flux (Jy  \kms):} 
~~~Shostak et al. (1984) & 1.3$\pm$0.3 & 1.7$\pm$0.2 &  10.5$\pm$1.0 &
--- & --- \\
~~~New VLA & 0.87$\pm$0.13 & 1.30$\pm$0.20 & 0.52$\pm$0.08 &
1.48$\pm$0.20 & 2.63$\pm$0.40 \\
$M_{HI}$ ($10^9 M_\odot$) & 1.5$\pm$0.2   &2.2$\pm$0.3 & 0.9$\pm$0.1 & 
2.5$\pm$0.4 & 4.0$\pm$0.6 \\
\enddata
\tablenotetext{\dagger} {Assuming distance of 85 Mpc for SQ.}

\end{deluxetable}

\clearpage
\begin{deluxetable}{lcccccc}
\tablecaption{Satellite Galaxies 
\label{table:sat}}
\tablehead{
\colhead{Anon.}         &
\colhead{size\tablenotemark{\dagger}}       & 
\colhead{Type}      &  
\colhead{$V_{\odot}$}       & 
\colhead{$\int S_vdv$}       &  
\colhead{$\Delta V_{20}$}       &  
\colhead{M$_{HI}$} \\
\colhead{No.}  &  
\colhead{(arcmin)}  &  \colhead{} &  
\colhead{(\kms)}  &  
\colhead{(Jy \kms)}   &
\colhead{(\kms)}  &  \colhead{($10^9M_\odot$)}  }
\startdata
2   &  $0.45 \times 0.20$&  
~Sc& 6369$\pm$10&  0.32$\pm$0.05&  86$\pm$10&  
0.54$\pm$0.08 \\
4   &  $0.35 \times 0.10$&  ~SB& 6057$\pm$10&  0.70$\pm$0.10&  129$\pm$10&
1.08$\pm$0.20 \\
6   &  $0.20 \times 0.10$&  ~S&  6047$\pm$10&  0.22$\pm$0.03&   65$\pm$10&
0.34$\pm$0.05 \\   
\enddata
\tablenotetext{\dagger} {Sizes were measured on the blue POSS print.}
\end{deluxetable}

\end{document}